# Thermodynamics and energy condition analysis for Van-der-Waals EOS without viscous cosmology.

**Alokananda Kar[1] Shouvik Sadhukhan[2] Surajit Chattopadhyay[3]**

[1]Department of Physics; University of Calcutta; 92 APC Road, Kolkata 700009, West Bengal, India

[2]Department of Physics; Indian Institute of Technology, Kharagpur 721302, West Bengal, India

[3]Department of Mathematics, Amity University Major Arterial Road, Action Area II, Rajarhat New Town, Kolkata 700135, India

[1]Email: alokanandakar@gmail.com, [2]Email : shouvikphysics1996@gmail.com, Correspondence Email : surajitchatto@outlook.com

**Abstract**

In this paper we have studied Van-Der-Waals fluid system with the generalized EOS as $p = w(\rho,t)\rho + f(\rho)$ as discussed in the recent works of Kremer, G.M [1, 4], Vardiashvili, G [2], Jantsch, R.C [3], Capozziello, S [5]. $w(\rho,t)$ and $f(\rho)$ are functions of energy density and time that are different for the three types of Van der Waals fluid which are one parameter model, two parameters model and three parameters model. We have studied the changes in the parameters for different cosmic phases. We have also investigated the thermodynamics and the stability conditions for these three models. Finally, we have resolved the finite time future singularity problems.

1. Introduction

After the discovery of cosmic acceleration various models of Dark energy have been proposed to explain the accelerated expansion of universe. Each of these models explain some specific element of acceleration. To explain this acceleration, we need various cosmological components and exotic equation of state. Just like Chaplygin gas model, Van-der-Waals EOS has received some attention to describe dark energy and standard matter as single fluid. We try to examine the ability of this model to reproduce the accelerated universe. In this paper we explore three different parametrizations of Van-der-Waal EOS as described in papers [1-6].

In all the Dark Energy models the basic target is to explain the existence of exotic matters that exerts the negative pressure necessary for accelerated expanding universe. Van-der-Waals fluid model is one of those models that explains the possible existence of negative pressure during cosmic expansion without considering any Dark Energy scalar field. Even without discussing the cause behind this Van-der-Waals nature of cosmic fluid in early universe, it has helped us to resolve a lot of cosmological problems classically. [6-20]

The well-known idea of cosmic inflation, Phantom era Finite time future singularity can be solved with this Van-der-Waals fluid model. Cosmic inflationary discussion has already been done by several authors. [6] We have tried to discuss most of the classical geometrical cosmic problems with the help of these models. From the thermodynamic concept of Van-der-Waals system we may construct three different form of Van-der-Waals fluid EOS in cosmology and they are One parameter model, Two parameters model and three parameters model. All those parameters carry different independent information about cosmic evolutionary phases like EOS parameter. These models also give us the time varying Scale factor, varying EOS parameters which can provide data from phantom era also. [21-40]

The basic motivation behind studying Van-der-Waals equation of state in this work is to examine the role of non-linear fluid in cosmology. Van-der-Waals model is the most generalized non-linear model. The non-linearity is introduced in the model by making the equation of state parameters $\omega$, density dependent. $\omega$ can be made local time dependent also but then it will violate least action principle. In that case we would need to introduce viscosity which is beyond the scope of this paper. However, we shall observe that $\omega$ evolves with changing scale factor of the universe which is a function of proper time. Hence these models obey least action principle. [41-62] [63, 64]

The paper is organized as follows: in section 2 we gave an overview of the Van-der-Waals fluid model and the evolution profile of the functions. In section 3 we discussed various thermodynamics quantities and also studied the stability of the models. In section 4 contains the overview of thermodynamics energy conditions and finite time future singularity problems. In section 5 we have investigated the energy conditions for the three models. Section 6 discusses the singularity problems and finally, section 7 gives the scalar field corresponding to our model.

2. **Overview of Nonlinear (Van-Der-Waals) model and its evolution profile**

The equation of state of Van-der-Waals system as discussed in the paper [1,4] is represented as equation (1).

$$p = w(\rho, t)\rho + f(\rho)$$

(1)

Where $w(\rho, t)$ and $f(\rho)$ are the two functions of $\rho$.

Considering $p = -\left(\frac{\partial U}{\partial V}\right)_S$ and $\rho = \frac{U}{V}$ (According to [12,13]) we may write the following differential equation for internal energy.

$$-\left(\frac{\partial U}{\partial V}\right)_S = w\left(\frac{U}{V}, t\right)\frac{U}{V} + f\left(\frac{U}{V}\right)$$

(2)

Here U = Internal energy and V = Volume scale factor = $(a(t))^3$, $a(t)$ = Scale factor of the Universe

We shall use equation (2) in our study in the following sections. Based on the choice of $w(\rho, t)$ and $f(\rho)$ we obtained three types of models, namely one parameter model, two parameters model and three parameters models as represented in table I as discussed in papers [1-6]. As previously mentioned, we have tried to introduce the nonlinear homogeneous model in the view of Van-Der-Waals EOS. Therefore, we try to simplify the models so that the nonlinearity in them is not be destroyed by keeping only the first order terms of denominators. We will simplify those models just to derive the internal energy as a function of scale factor.

**Table I**

| Models and their simplifications | | |
|---|---|---|
| One parameter model | Two parameters model | Three parameters model |
| $p = \dfrac{8w\rho}{3-\rho} - 3\rho^2$ | $p = \dfrac{\gamma\rho}{1 - \dfrac{1}{3\rho_c}\rho} - \dfrac{9\gamma}{8\rho_c}\rho^2$ | $p = \dfrac{\gamma\rho}{1-\beta\rho} - \alpha\rho^2$ |
| One parameter model with our assumption | Two parameters model with our assumption | Three parameters model with our assumption |
| $p = \dfrac{8w}{3}\rho + \left(\dfrac{8w}{9} - 3\right)\rho^2$ | $p = \gamma\rho - \dfrac{19\gamma}{24\rho_c}\rho^2$ | $p = \gamma\rho + (\gamma\beta - \alpha)\rho^2$ |

Using equation (2) and the EOS of table I we obtained various functional forms, such as, the pressures, energy densities, EOS parameters and Internal energies with following functional formats.

### 2.1. One Parameter model

Here we are providing the internal energy for one parameter model which can be derived from the differential equation for the same. [Appendix]

$$U = \frac{(8w+3)V}{(27-8w)+(9+24w)U_0 V^{\frac{8w+3}{3}}}$$

(3)

The energy density can be found from equation (3) easily as follows.

$$\rho = \frac{(8w+3)}{(27-8w)+(9+24w)U_0 V^{\frac{8w+3}{3}}}$$

(4)

And the pressure will be as follows,

$$p = \frac{8w\left(\frac{(8w+3)}{(27-8w)+(9+24w)U_0 V^{\frac{8w+3}{3}}}\right)}{3-\left(\frac{(8w+3)}{(27-8w)+(9+24w)U_0 V^{\frac{8w+3}{3}}}\right)} - 3\left(\frac{(8w+3)}{(27-8w)+(9+24w)U_0 V^{\frac{8w+3}{3}}}\right)^2$$

(5)

Thus, we can get the variable pressure and density and therefore the EOS parameter which is as follows.

$$w_{EOS} = \frac{P}{\rho} = \frac{8w}{3-\left(\frac{(8w+3)}{(27-8w)+(9+24w)U_0 V^{\frac{8w+3}{3}}}\right)} - 3\left(\frac{(8w+3)}{(27-8w)+(9+24w)U_0 V^{\frac{8w+3}{3}}}\right)$$

(6)

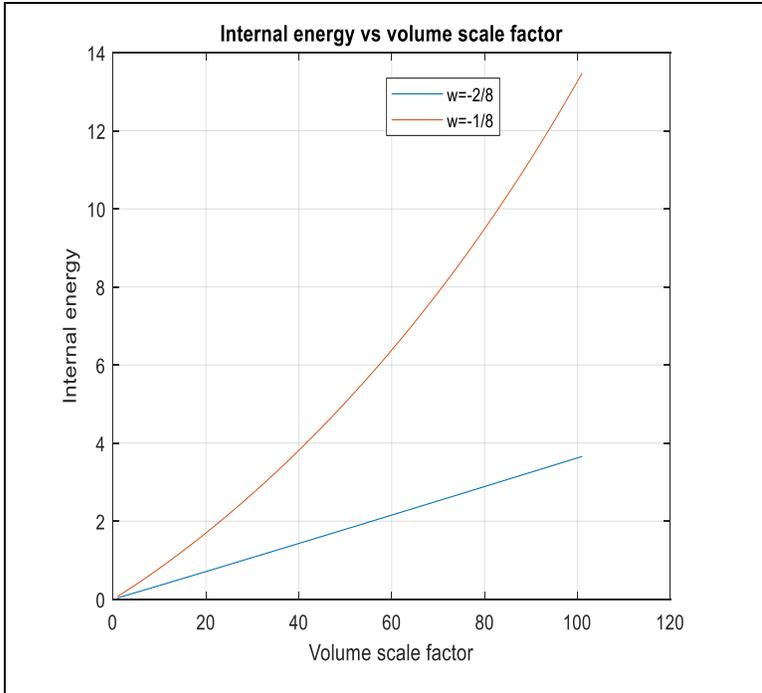

Figure 1: Evolution of Internal Energy with volume scale factor

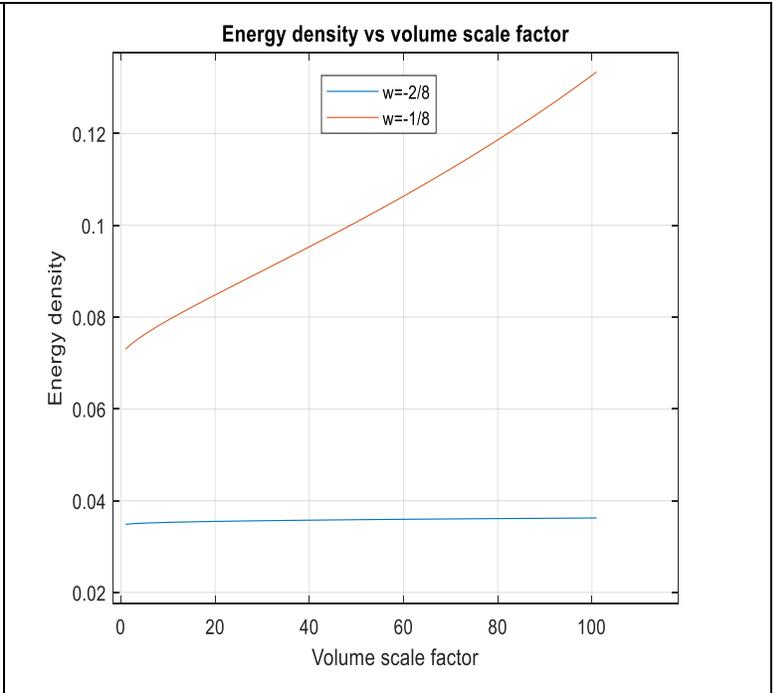

Figure 2: Evolution of Energy Density with volume scale factor

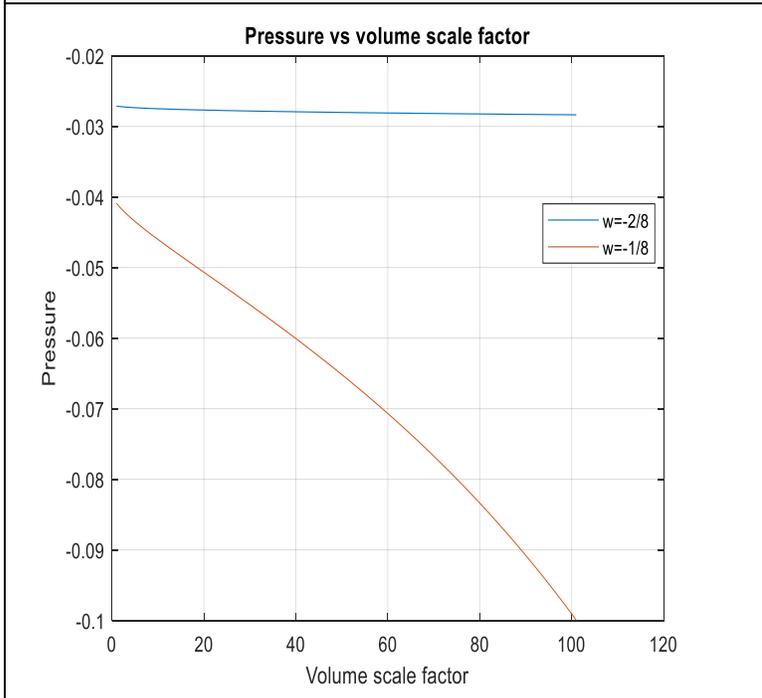

Figure 3: Evolution of fluid pressure with volume scale factor

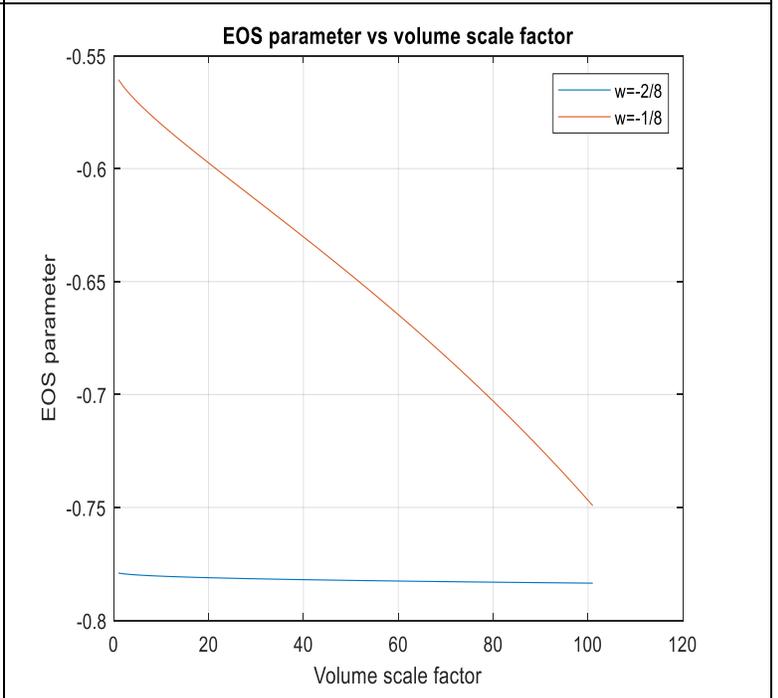

Figure 4 : Evolution of EOS Parameter with volume scale factor

## 2.2. Two Parameters model

Here we are providing the internal energy for two parameters model which can be derived from the differential equation for the same. [Appendix]

$$U = \frac{24(\gamma + 1)\rho_c V}{19\gamma + 24U_0\rho_c(1 + \gamma)V^{1+\gamma}}$$

(7)

The energy density can be found from equation (7) easily as follows.

$$\rho = \frac{24(\gamma + 1)\rho_c}{19\gamma + 24U_0\rho_c(1 + \gamma)V^{1+\gamma}}$$

(8)

And the pressure will be as follows,

$$p = \frac{\gamma\left(\frac{24(\gamma + 1)\rho_c}{19\gamma + 24U_0\rho_c(1 + \gamma)V^{1+\gamma}}\right)}{1 - \frac{1}{3\rho_c}\left(\frac{24(\gamma + 1)\rho_c}{19\gamma + 24U_0\rho_c(1 + \gamma)V^{1+\gamma}}\right)} - \frac{9\gamma}{8\rho_c}\left(\frac{24(\gamma + 1)\rho_c}{19\gamma + 24U_0\rho_c(1 + \gamma)V^{1+\gamma}}\right)^2$$

(9)

Thus, we can get the variable pressure and density and therefore the EOS parameter which is as follows.

$$w_{EOS} = \frac{P}{\rho} = \frac{\gamma}{1 - \frac{1}{3\rho_c}\left(\frac{24(\gamma + 1)\rho_c}{19\gamma + 24U_0\rho_c(1 + \gamma)V^{1+\gamma}}\right)} - \frac{9\gamma}{8\rho_c}\left(\frac{24(\gamma + 1)\rho_c}{19\gamma + 24U_0\rho_c(1 + \gamma)V^{1+\gamma}}\right)$$

(10)

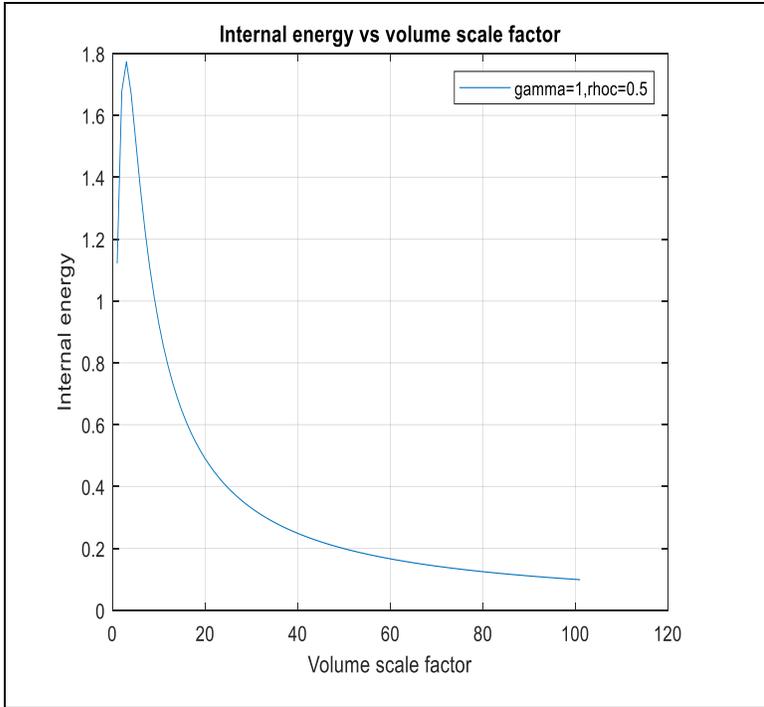

Figure 5: Evolution of Internal Energy with volume scale factor

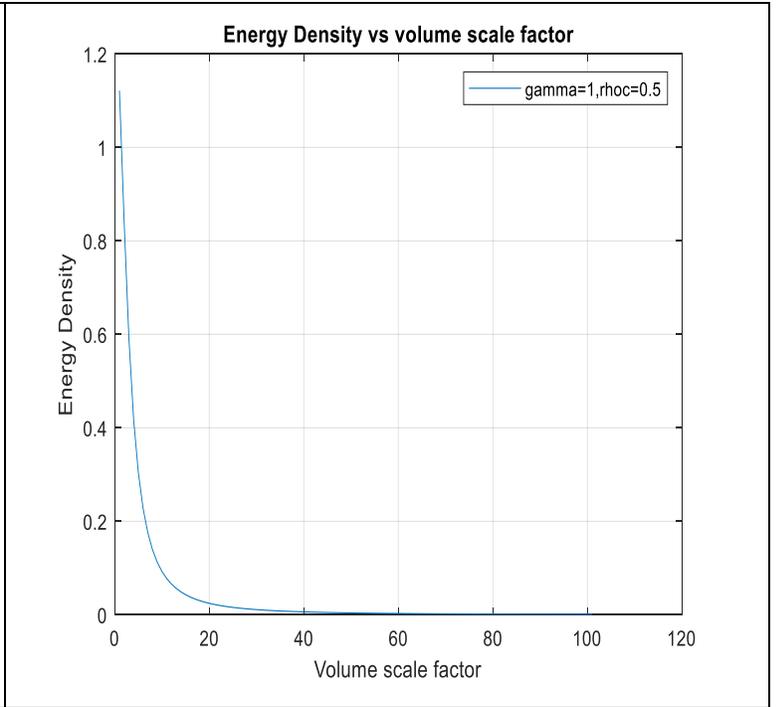

Figure 6: Evolution of Energy Density with volume scale factor

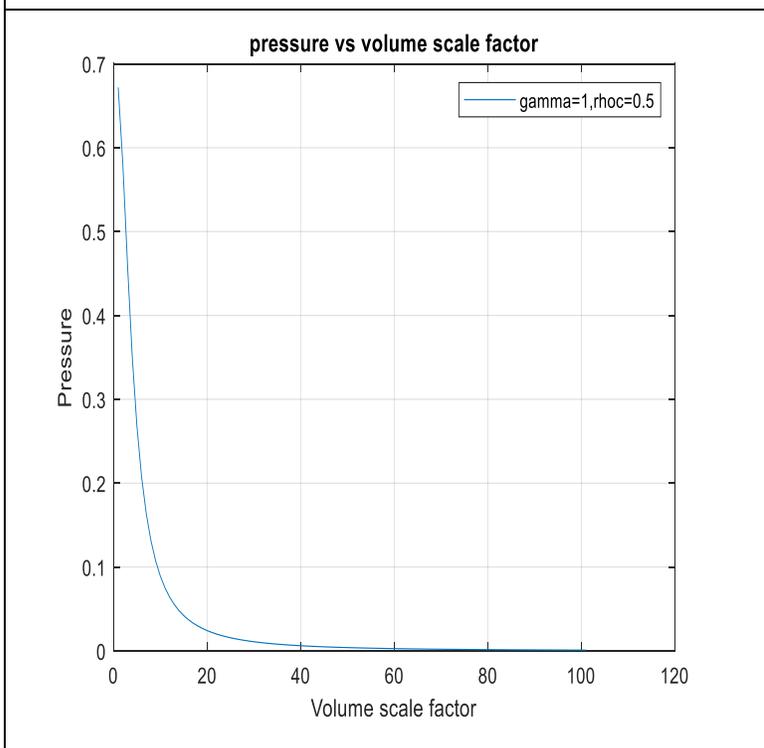

Figure 7: Evolution Pressure with volume scale factor

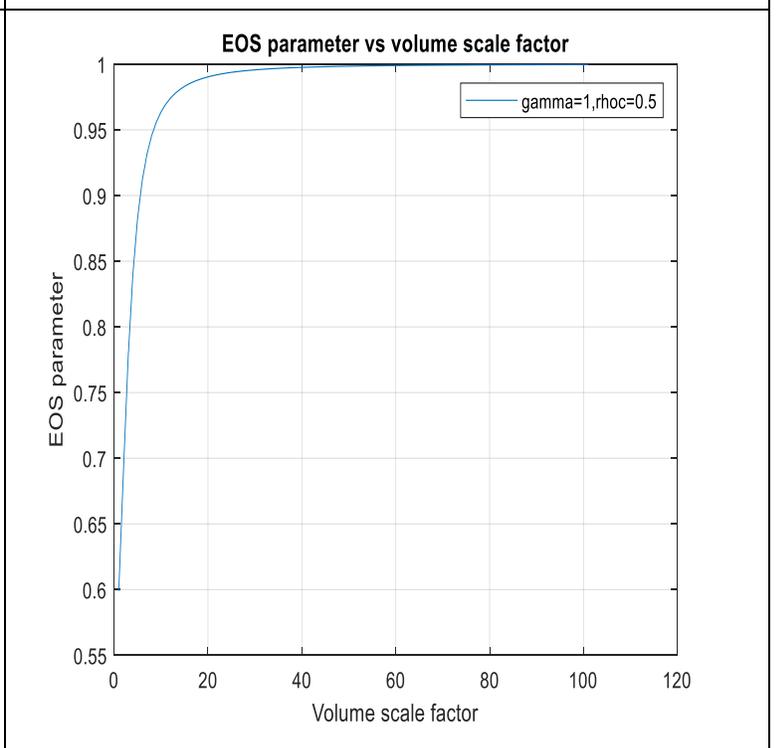

Figure 8: Evolution EOS Parameter with volume scale factor

### 2.3. Three Parameters model

Here we are providing the internal energy for three parameters model which can be derived from the differential equation for the same. [Appendix]

$$U = \frac{(\gamma + 1)V}{(\alpha - \beta\gamma) + U_0(1 + \gamma)V^{1+\gamma}}$$

(11)

The energy density can be found from equation (11) easily as follows.

$$\rho = \frac{(\gamma + 1)}{(\alpha - \beta\gamma) + U_0(1 + \gamma)V^{1+\gamma}}$$

(12)

And the pressure will be as follows,

$$p = \frac{\gamma\left(\frac{(\gamma + 1)}{(\alpha - \beta\gamma) + U_0(1 + \gamma)V^{1+\gamma}}\right)}{1 - \beta\left(\frac{(\gamma + 1)}{(\alpha - \beta\gamma) + U_0(1 + \gamma)V^{1+\gamma}}\right)} - \alpha\left(\frac{(\gamma + 1)}{(\alpha - \beta\gamma) + U_0(1 + \gamma)V^{1+\gamma}}\right)^2$$

(13)

Thus, we can get the variable pressure and density and therefore the EOS parameter which is as follows.

$$w_{EOS} = \frac{P}{\rho} = \frac{\gamma}{1 - \beta\left(\frac{(\gamma + 1)}{(\alpha - \beta\gamma) + U_0(1 + \gamma)V^{1+\gamma}}\right)} - \alpha\left(\frac{(\gamma + 1)}{(\alpha - \beta\gamma) + U_0(1 + \gamma)V^{1+\gamma}}\right)$$

(14)

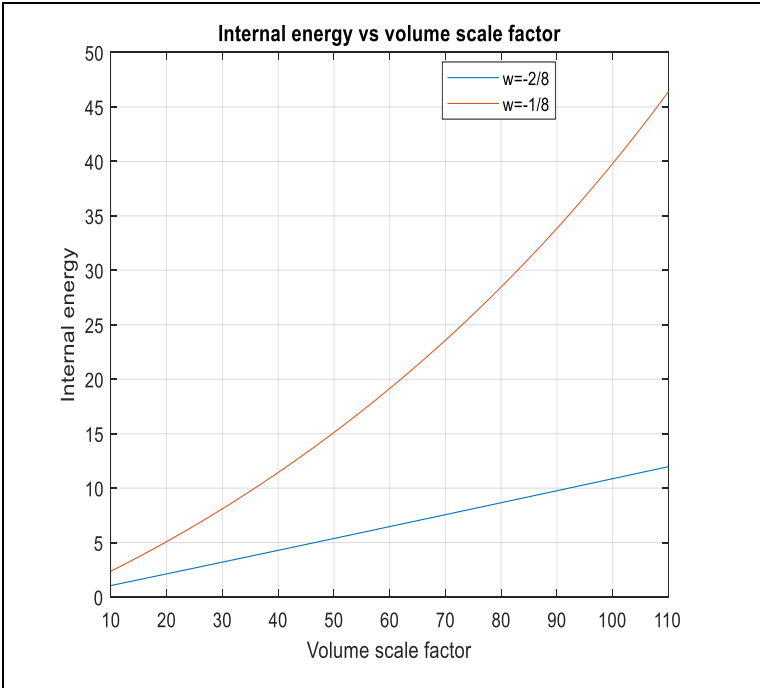

Figure 9: Evolution Internal Energy with volume scale factor

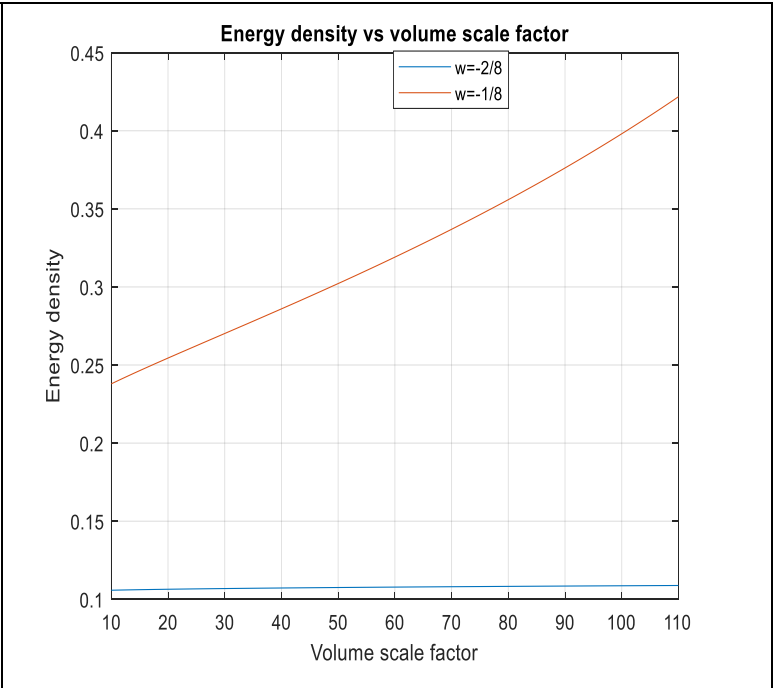

Figure 10: Evolution of Energy Density with volume scale factor

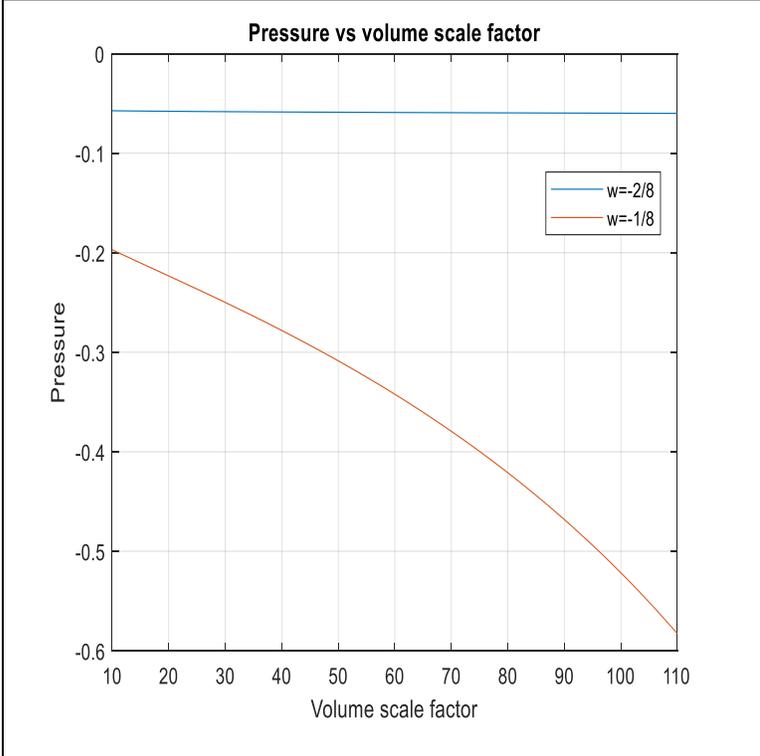

Figure 11: Evolution of Pressure with volume scale factor

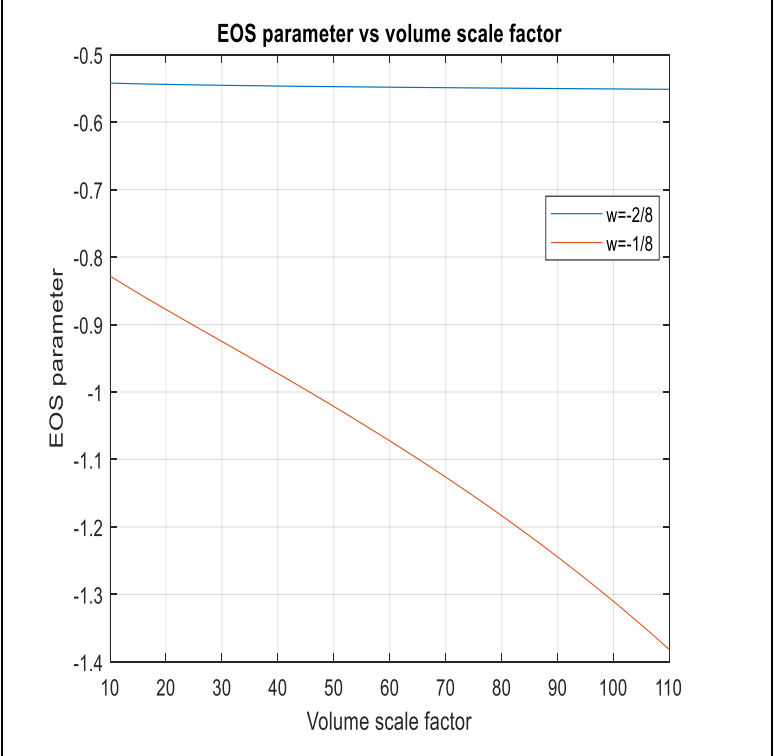

Figure 12: Evolution of EOS Parameter with volume scale factor

We have used the values of $\alpha, \beta, \gamma$ as 3, 1/3, -1/3 to -2/3 respectively. Here we have drawn a correspondence between three and one parameter model to obtain the value of $\alpha, \beta\ and\ \gamma$. Within this limit we get negative pressure which is a necessary condition for accelerating expanding universe. (Fig. 1-12)

## 3. Overview of Cosmic Thermodynamics and its analysis with our models with the discussion on thermodynamics stability

In this section we have discussed the thermodynamic properties of the system through its internal entropy. We derived the change of differential entropy through dimensional analysis as done in papers [12,13]. The temperature can be derived by the following equation [14].

$$\frac{\dot{T}}{T} = -3\frac{\dot{a}}{a}\frac{\partial p}{\partial \rho} = -\frac{\dot{V}}{V}\frac{\partial p}{\partial \rho}$$

(15)

We know that $T = \left(\frac{\partial U}{\partial S}\right)_V$ and from dimensional analysis technique we have $[U] = [T][\Delta S]$. [12-13] We obtained the results that are represented in table II. [12,13]

**Table II**

| Evolution of Temperature | |
|---|---|
| One Parameter Model | $T = \exp\left(-\int_{V_0}^{V}\frac{dV}{V}\left[\frac{24w}{(3-\rho)^2} - 6\rho\right]\right)$ |
| Two Parameters Model | $T = \exp\left(-\int_{V_0}^{V}\frac{dV}{V}\left[\frac{\gamma}{\left(3-\frac{1}{3\rho_c}\rho\right)^2} - \frac{9\gamma}{4\rho_c}\rho\right]\right)$ |
| Three Parameters Model | $T = \exp\left(-\int_{V_0}^{V}\frac{dV}{V}\left[\frac{\gamma}{(1-\beta\rho)^2} - 2\alpha\rho\right]\right)$ |
| **Evolution of $\Delta S$ where $S$ is the entropy** | |
| One Parameter Model | $\Delta S = \dfrac{(8w+3)(9+24w)V^{\frac{8w+6}{3}}}{\left[(27-8w)+(9+24w)V^{\frac{8w+3}{3}}U_0\right]^2}\dfrac{U_0}{\exp\left(-\int_{V_0}^{V}\frac{dV}{V}\left[\frac{24w}{(3-\rho)^2}-6\rho\right]\right)}$ |

| Two Parameters Model | $\Delta S = \dfrac{[24(\gamma+1)]^2 V^{\gamma+2}\rho_c}{[19\gamma + 24(1+\gamma)\rho_c V^{\gamma+1} U_0]^2} \dfrac{U_0}{\exp\left(-\int_{V_0}^{V}\dfrac{dV}{V}\left[\dfrac{\gamma}{\left(3-\dfrac{1}{3\rho_c}\rho\right)^2} - \dfrac{9\gamma}{4\rho_c}\rho\right]\right)}$ |
|---|---|
| Three Parameters Model | $\Delta S = \dfrac{(\gamma+1)^2 V^{\gamma+2}}{[(\alpha-\beta\gamma)+(\gamma+1)V^{\gamma+1}U_0]^2} \dfrac{U_0}{\exp\left(-\int_{V_0}^{V}\dfrac{dV}{V}\left[\dfrac{\gamma}{(1-\beta\rho)^2} - 2\alpha\rho\right]\right)}$ |

The differential entropy discussed in the table II verifies the validity of second law of thermodynamics in our model. For the condition $U_0 > 0$ we can have $\Delta S > 0$. Therefore, we can conclude that all of our models obey the second law of thermodynamics.

Now we can discuss the thermodynamics stability of those three models with respect to expanding universe. As we know that the expanding universe fluid pressure and internal energy should satisfy the following three rules of model stability. [32]

Rule I: $\left(\dfrac{\partial p}{\partial V}\right)_S < 0$

Rule II: $\left(\dfrac{\partial p}{\partial V}\right)_T < 0$

Rule III: $c_V = T\left(\dfrac{\partial S}{\partial T}\right)_V > 0$

From previously derived expression of pressure, density and entropy, we investigated the stability conditions for the above three models and the results are represented in table III.

**Table III**

| Discussion on stability conditions | | |
|---|---|---|
| One Parameter model | Two Parameters model | Three Parameters model |
| $\left(\dfrac{\partial p}{\partial V}\right)_S < 0$ | $\left(\dfrac{\partial p}{\partial V}\right)_S < 0$ | $\left(\dfrac{\partial p}{\partial V}\right)_S < 0$ |
| $\left(\dfrac{\partial p}{\partial V}\right)_T < 0$ | $\left(\dfrac{\partial p}{\partial V}\right)_T < 0$ | $\left(\dfrac{\partial p}{\partial V}\right)_T < 0$ |
| $c_V = T\left(\dfrac{\partial S}{\partial T}\right)_V < 0$ | $c_V = T\left(\dfrac{\partial S}{\partial T}\right)_V < 0$ | $c_V = T\left(\dfrac{\partial S}{\partial T}\right)_V < 0$ |
| For $U_0 > 0$ or $(8w+3)(9+24w) > 0$ | For $U_0 > 0$ or $(24(\gamma+1))^2 \rho_c > 0$ | For $U_0 > 0$ or $(\gamma+1)^2 > 0$ |

From the table III we can conclude that all those models are stable w.r.t. the first two conditions. But with the analysis of specific heat we can't stabilize these models.

## 4. Overview of Thermodynamics Energy conditions and Finite time future singularity problems

For finite time singularity problems, we know the following types; [57] [50]

- Type I (known as Big Rip): for $t = t_s$ we have $a \to \infty$; $\rho \to \infty$; $p \to \infty$
- Type II (known as Sudden Singularity): For $t = t_s$ we have $a \to a_s$; $\rho \to \rho_s$; $p \to \infty$
- Type III: For $t = t_s$ we have $a \to a_s$; $\rho \to \infty$; $p \to \infty$ ; it happens only in the EOS type of $p = -\rho - A\rho^\alpha$.
- Type IV: For $t = t_s$ we have $a \to a_s$; $\rho \to \rho_s$; $p \to p_s$ This kind of singularity mainly comes into play when $p = -\rho - f(\rho)$.

From the Raychaudhuri's equation with a congruence of time-like and null-like geodesics we get,

$$\frac{d\theta}{d\tau} = -\frac{1}{3}\theta^2 - \sigma_{\mu\nu}\sigma^{\mu\nu} + \omega_{\mu\nu}\omega^{\mu\nu} - R_{\mu\nu}u^\mu u^\nu \tag{16}$$

And

$$\frac{d\theta}{d\tau} = -\frac{1}{3}\theta^2 - \sigma_{\mu\nu}\sigma^{\mu\nu} + \omega_{\mu\nu}\omega^{\mu\nu} - R_{\mu\nu}n^\mu n^\nu \tag{17}$$

Where $\theta$ is the expansion factor, $n^\mu n^\nu$ is the null vector, and $\sigma_{\mu\nu}\sigma^{\mu\nu}$ and $\omega_{\mu\nu}\omega^{\mu\nu}$ are, respectively, the shear and the rotation associated with the vector field $u^\mu u^\nu$. For attractive gravity we'll have the followings;

$$R_{\mu\nu}u^\mu u^\nu \geq 0 \text{ and } R_{\mu\nu}n^\mu n^\nu \geq 0$$

So, for our matter-fluid distribution we may write this condition as follows;

- NEC (Null Energy Condition) = $\rho + p \geq 0$
- WEC (Weak Energy Condition) = $\rho \geq 0$ and $\rho + p \geq 0$
- SEC (Strong Energy Condition) = $\rho + 3p \geq 0$ and $\rho + p \geq 0$
- DEC (Dominant Energy Condition) = $\rho \geq 0$ and $-\rho \leq p \leq \rho$

## 5. Cosmological energy conditions analysis with our models and their evolutions

We are avoiding the functional representations. Those have been given in appendix.

### 5.1. Representation of energy conditions for one parameter model

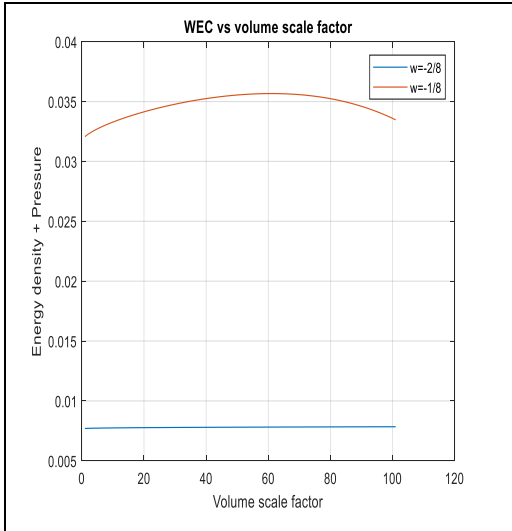

Figure 13: Pictorial representation of WEC

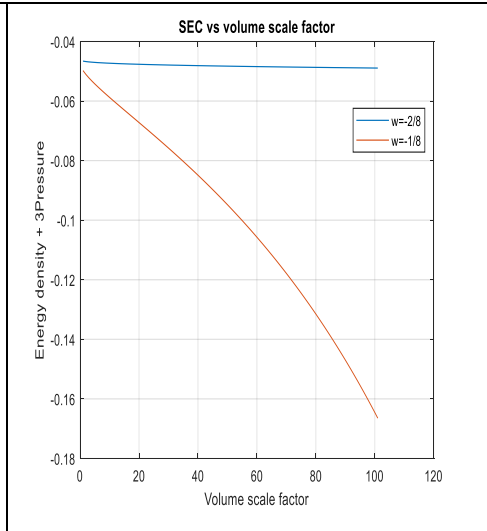

Figure 14: Pictorial representation of SEC

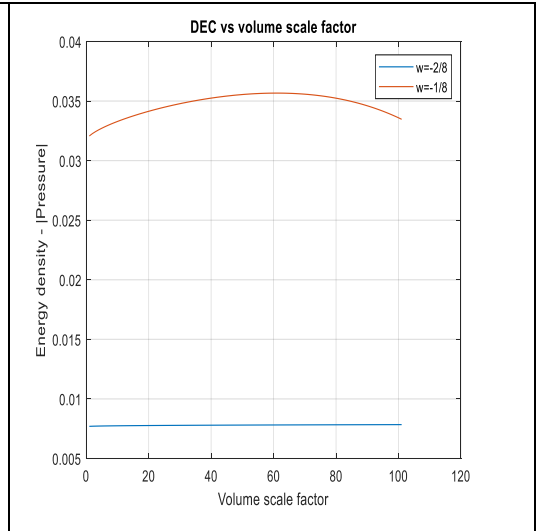

Figure 15: Pictorial representation of DEC

### 5.2. Representations of energy conditions for two parameters model

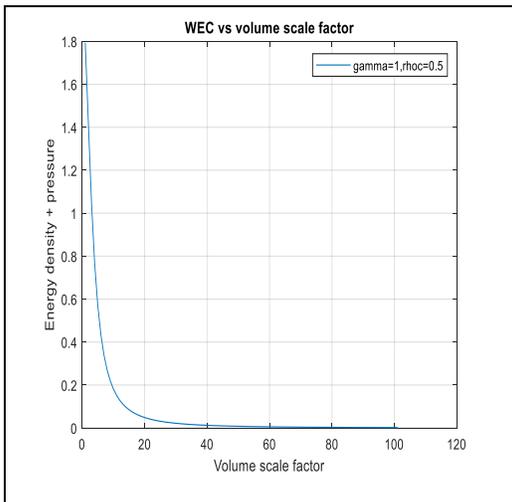

Figure 16: Pictorial representation of WEC

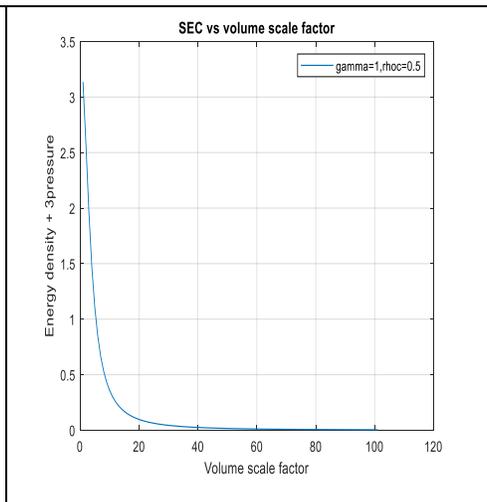

Figure 17: Pictorial representation of SEC

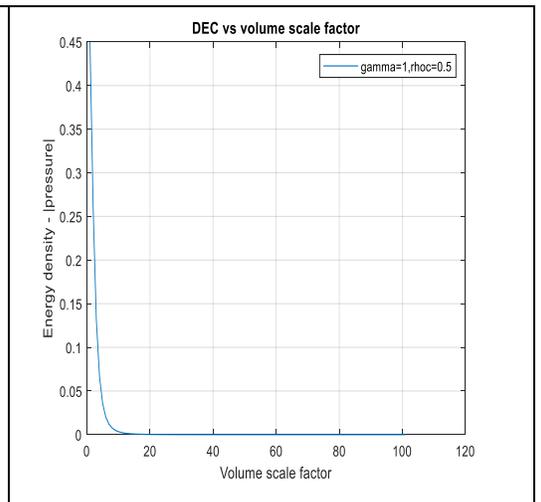

Figure 18: Pictorial representation of DEC

### 5.3. Representations of energy conditions for three parameters model

| 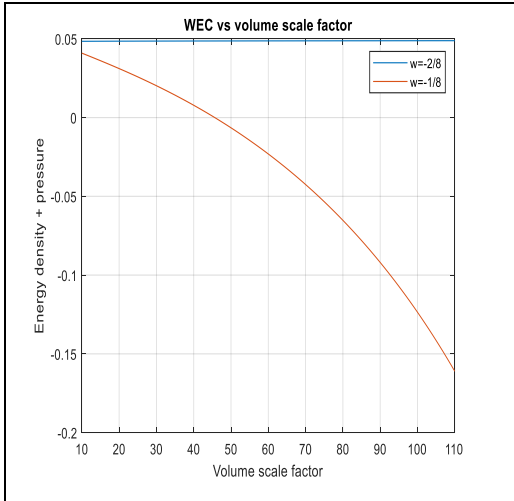 | 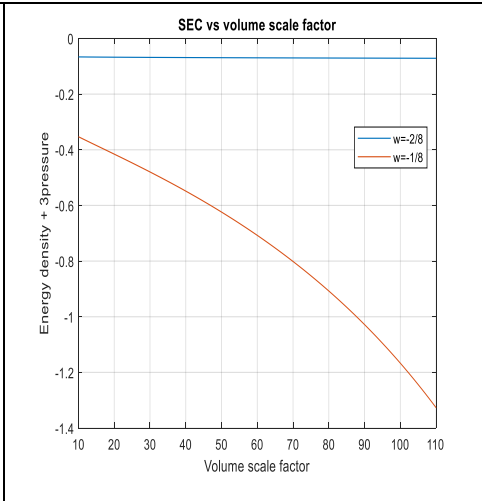 | 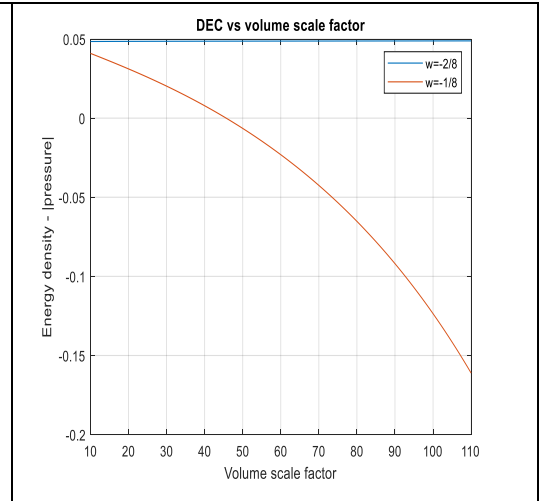 |
|---|---|---|
| Figure 19: Pictorial representation of WEC | Figure 20: Pictorial representation of SEC | Figure 21: Pictorial representation of DEC |

Therefore, we can observe that the one parameter and three parameters models satisfy all energy conditions except SEC where as two parameters model follows all those conditions for attractive gravity. Those conditions have been discussed with the previously considered values of constants. (Fig. 13-21)

## 6. Discussion of singularity problems

Here in this section we have discussed the resolution of finite time future singularity problems. We'll hereby consider a minimal coupling between Quintessence scalar field with Van-Der-Waals fluid and that's why we may write as follows. [51,52]

$$3H^2 = \rho_\phi + \rho$$

(18)

Where $\rho_\phi = scalar\ field\ density\ and\ \rho = Van-der-Waals\ fluid\ density$

Now from quintessence Klein Gordon equation with potential as a function of kinetic energy we can solve the scalar field energy density as follows.

$$\rho_\phi = mV^n$$

(19)

And

The generalized form of fluid density for all those three models can be written as follows.

$$\rho = \frac{A}{B + CV^D}$$

(20)

Here A, B, C and D are the constants that depend upon the parameters of those three models.

So, from the above three relations we may write as follows.

$$\int \frac{dV}{V} \left[ \frac{B + CV^D}{\frac{m}{3} BV^n + \frac{m}{3} CV^{n+D} + \frac{A}{3}} \right] = t + C_1$$

(21)

Where $C_1 = constant\ of\ integration \to +ve$

Here we have derived the scale factor vs time relation from the minimally coupled Friedmann equation. For $t \to 0$ we can't get any singularity on RHS of the equation (68). So, the initial singularity problem has been resolved classically. Now we got the constant of integration as a positive valued constant. So, we can't find any time $t \to t_s$ such that we get singularity in RHS of equation (68). Therefore, the finite time future singularity problems have also been resolved in phantom era.

## 7. Scalar field theory corresponding to the three models of Van-der-Waals EOS

Here in this section we'll show the corresponding scalar field and potential variation for Van-Der-Waals fluid cosmology. We'll use the general scalar field theory or Quintessence to find the scale factor variation of scalar field and its potential. So, the calculations are as follows. [67]

We know from Quintessence scalar field theory,

$$\rho = \frac{1}{2}\dot{\phi}^2 + V(\phi)\ and\ p = \frac{1}{2}\dot{\phi}^2 - V(\phi)$$

(22)

So, we get the kinetic term of scalar field theory and potential as follows.

$$\dot{\phi} = (p + \rho)^{\frac{1}{2}}$$

And

$$V(\phi) = \frac{1}{2}(\rho - p)$$

(23)

(24a)

$$KE = \frac{1}{2}\dot{\phi}^2 = \frac{1}{2}(p + \rho)$$

(24b)

Now this potential and kinetic term will get the form for the models as follows.

### 7.1. One parameter model:

From the derivations of Van-Der-Waals fluid energy density and pressure for one parameter model we can write as follows.

$$\dot{\phi} = \left( \frac{8w\left(\frac{(8w+3)}{(27-8w)+(9+24w)U_0 V^{\frac{8w+3}{3}}}\right)}{3 - \left(\frac{(8w+3)}{(27-8w)+(9+24w)U_0 V^{\frac{8w+3}{3}}}\right)} - 3\left(\frac{(8w+3)}{(27-8w)+(9+24w)U_0 V^{\frac{8w+3}{3}}}\right)^2 + \frac{(8w+3)}{(27-8w)+(9+24w)U_0 V^{\frac{8w+3}{3}}} \right)^{\frac{1}{2}}$$

(25)

And

$$V(\phi) = \frac{1}{2}\left( \frac{(8w+3)}{(27-8w)+(9+24w)U_0 V^{\frac{8w+3}{3}}} - \frac{8w\left(\frac{(8w+3)}{(27-8w)+(9+24w)U_0 V^{\frac{8w+3}{3}}}\right)}{3 - \left(\frac{(8w+3)}{(27-8w)+(9+24w)U_0 V^{\frac{8w+3}{3}}}\right)} + 3\left(\frac{(8w+3)}{(27-8w)+(9+24w)U_0 V^{\frac{8w+3}{3}}}\right)^2 \right)$$

(26)

So, the scale factor variation of those variables can be plotted as follows.

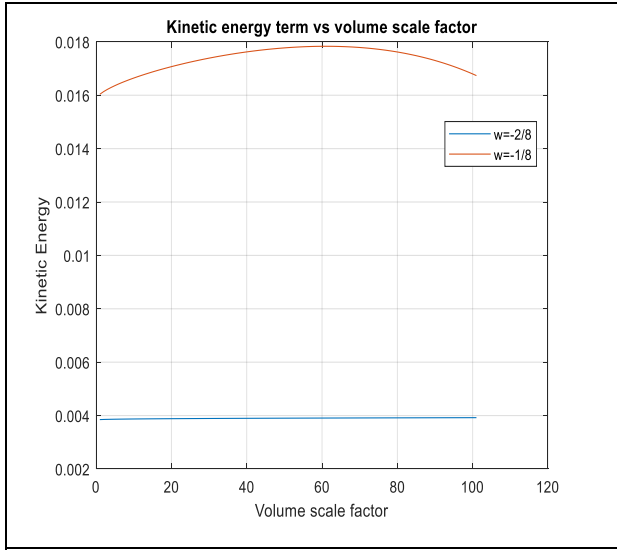

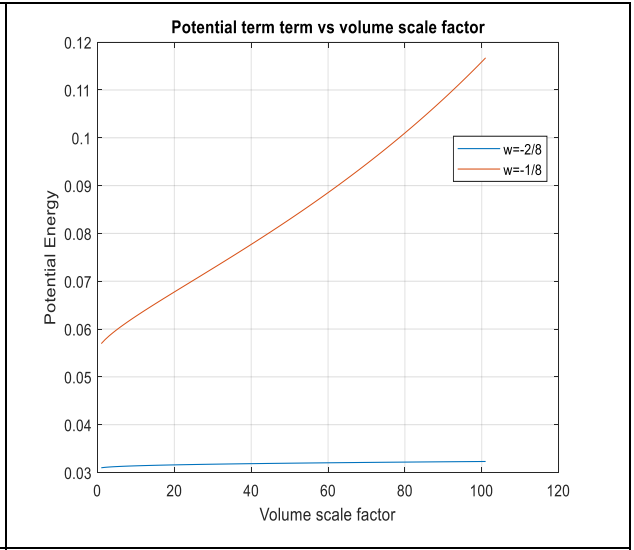

| Figure 22: Pictorial representation of evolution of Scalar field kinetic energy | Figure 23: Pictorial representation of evolution of scalar field potential energy |
|---|---|

### 7.2. Two parameters model:

From the derivations of Van-Der-Waals fluid energy density and pressure for one parameter model we can write as follows.

$$\dot{\phi} = \left( \frac{\gamma\left(\frac{24(\gamma+1)\rho_c}{19\gamma+24U_0\rho_c(1+\gamma)V^{1+\gamma}}\right)}{1-\frac{1}{3\rho_c}\left(\frac{24(\gamma+1)\rho_c}{19\gamma+24U_0\rho_c(1+\gamma)V^{1+\gamma}}\right)} - \frac{9\gamma}{8\rho_c}\left(\frac{24(\gamma+1)\rho_c}{19\gamma+24U_0\rho_c(1+\gamma)V^{1+\gamma}}\right)^2 + \frac{24(\gamma+1)\rho_c}{19\gamma+24U_0\rho_c(1+\gamma)V^{1+\gamma}} \right)^{\frac{1}{2}} \qquad (27)$$

And

$$V(\phi) = \frac{1}{2}\left( \frac{24(\gamma+1)\rho_c}{19\gamma+24U_0\rho_c(1+\gamma)V^{1+\gamma}} - \frac{\gamma\left(\frac{24(\gamma+1)\rho_c}{19\gamma+24U_0\rho_c(1+\gamma)V^{1+\gamma}}\right)}{1-\frac{1}{3\rho_c}\left(\frac{24(\gamma+1)\rho_c}{19\gamma+24U_0\rho_c(1+\gamma)V^{1+\gamma}}\right)} + \frac{9\gamma}{8\rho_c}\left(\frac{24(\gamma+1)\rho_c}{19\gamma+24U_0\rho_c(1+\gamma)V^{1+\gamma}}\right)^2 \right) \qquad (28)$$

So, the scale factor variation of those variables can be plotted as follows.

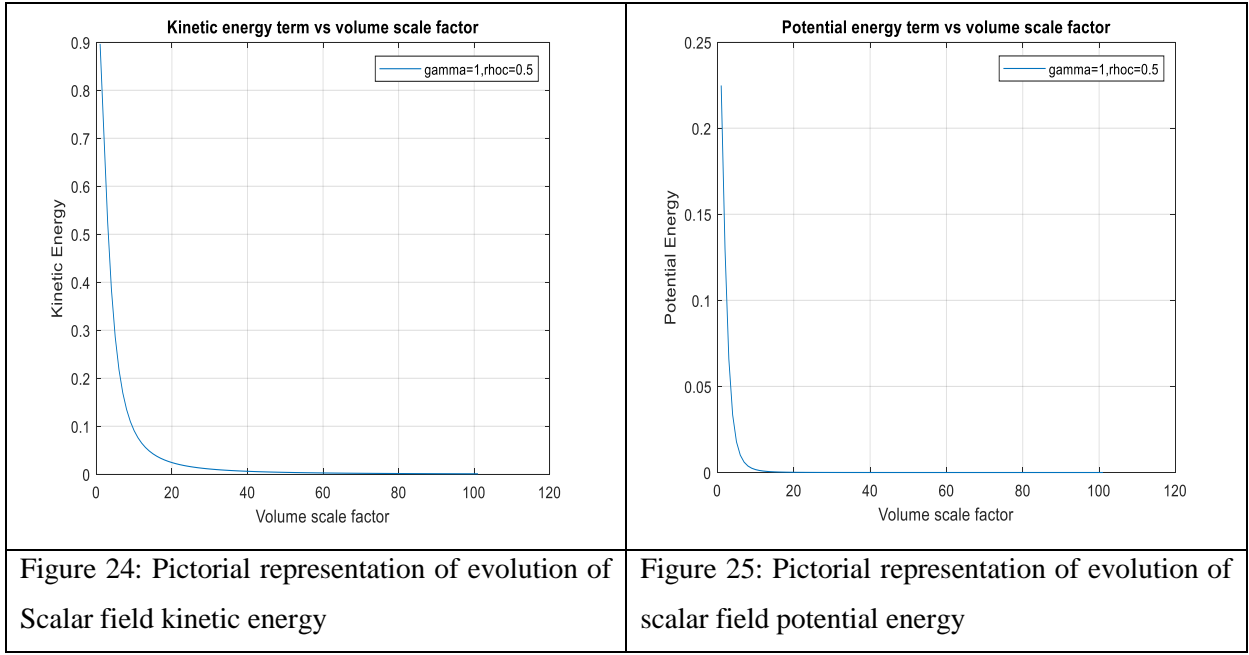

| Figure 24: Pictorial representation of evolution of Scalar field kinetic energy | Figure 25: Pictorial representation of evolution of scalar field potential energy |

### 7.3. Three parameters model:

From the derivations of Van-Der-Waals fluid energy density and pressure for one parameter model we can write as follows.

$$\dot{\phi} = \left( \frac{\gamma\left(\frac{(\gamma+1)}{(\alpha-\beta\gamma)+U_0(1+\gamma)V^{1+\gamma}}\right)}{1-\beta\left(\frac{(\gamma+1)}{(\alpha-\beta\gamma)+U_0(1+\gamma)V^{1+\gamma}}\right)} - \alpha\left(\frac{(\gamma+1)}{(\alpha-\beta\gamma)+U_0(1+\gamma)V^{1+\gamma}}\right)^2 + \frac{(\gamma+1)}{(\alpha-\beta\gamma)+U_0(1+\gamma)V^{1+\gamma}} \right)^{\frac{1}{2}} \quad (29)$$

And

$$V(\phi) = \frac{1}{2}\left( \frac{(\gamma+1)}{(\alpha-\beta\gamma)+U_0(1+\gamma)V^{1+\gamma}} - \frac{\gamma\left(\frac{(\gamma+1)}{(\alpha-\beta\gamma)+U_0(1+\gamma)V^{1+\gamma}}\right)}{1-\beta\left(\frac{(\gamma+1)}{(\alpha-\beta\gamma)+U_0(1+\gamma)V^{1+\gamma}}\right)} + \alpha\left(\frac{(\gamma+1)}{(\alpha-\beta\gamma)+U_0(1+\gamma)V^{1+\gamma}}\right)^2 \right) \quad (30)$$

So, the scale factor variation of those variables can be plotted as follows.

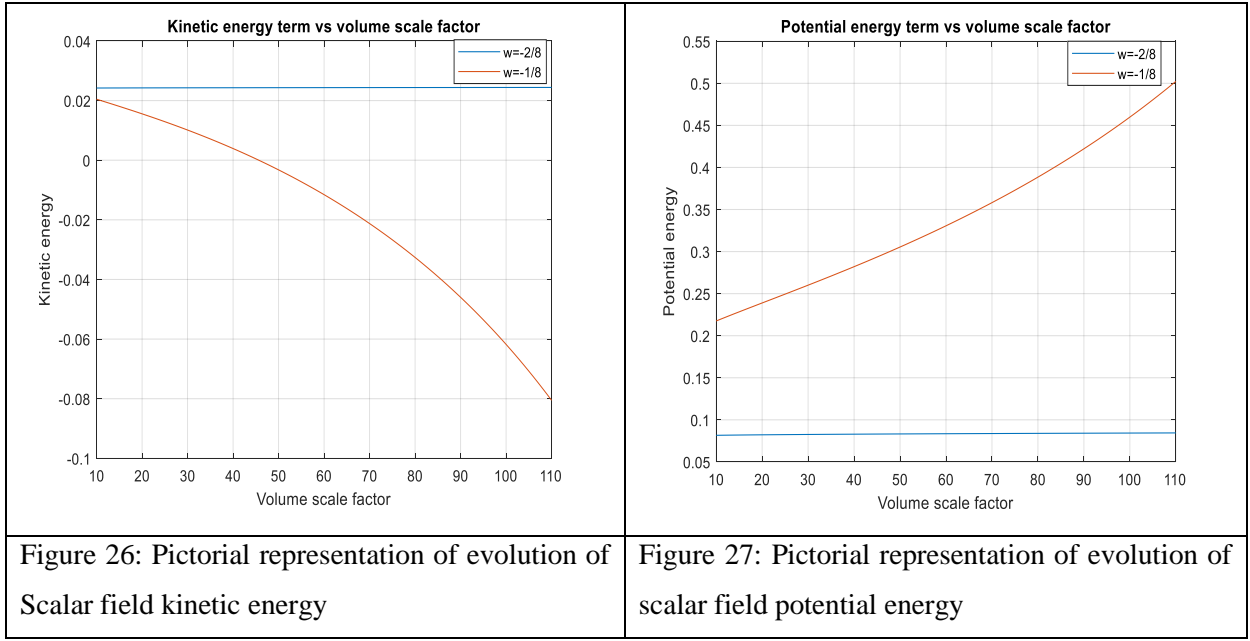

| Figure 26: Pictorial representation of evolution of Scalar field kinetic energy | Figure 27: Pictorial representation of evolution of scalar field potential energy |

For one parameter model we can see kinetic energy is less than potential energy. In case of three parameters model the K.E term will show negative result for w = -1/8 in late time accelerated universe. So, we have to skip this value of w and consider it to be between -2/8 and -1/8. Both the one and three parameters model satisfy the slow-roll model of inflation i.e. K.E < P.E. Two parameters model doesn't satisfy the mentioned criteria of inflation. (Fig. 22-27)

## 8. Results and comments

This work contained discussion on non-linear van-der-Waals type fluid cosmic system. We have discussed three types of nonlinear models depending upon the parameters used in the EOS.

From the pictorial discussion of these three models for pressures, energy densities and internal energies we can reach some important results and conclusions. For the one parameter and three parameters model we have found negative pressures. The energy densities and internal energies of these two models also give positive values. These trends of evolution of pressures and energy densities support the accelerating expansion of universe. The equation of state parameter represents the acceleration on post inflationary stage of universe. We have also represented a smooth transition between pre-inflationary phantom phase and post inflationary Quintessence phase. Both phases and their evolutionary conditions can be represented with our nonlinear models. The two parameters model represents positive pressures and its evolution with positive energy density. Therefore, the two parameters model can be useful to represent the attractive or contracting phases of universe which should come at the point of graceful exit after cosmic inflation. The adequate to inadequate inflation transition can also be explained with such equation of state. All these results are represented with the figures 1 to 12.

From the stability conditions discussion, we can observe that the models became unstable for specific heat at constant volume. Therefore, some additional term should be introduced to make it stable. It can be done by introducing a minimally coupled scalar field and discussing generalized thermodynamics with that coupled scalar field which is beyond the scope of this paper.

From discussion of energy conditions, we can observe that the one parameter and three parameters models satisfy all energy conditions except SEC where as two parameters model follows all the conditions for attractive gravity. These conditions have been discussed with the previously considered values of constants.

From minimally coupled field analysis we have reconstructed a scale factor that can remove all kind of finite time future singularities. It can also provide the resolution of initial singularity i.e. at $t \to 0$ we get $a \to a_{finite}$. As previously discussed that our model can explain the phantom phase properly in terms of universe expansion, it can also provide a smooth transition from phantom phase to quintessence phase. It has also provided finite dimension at the end of phantom phase and removed finite time future singularities. Therefore, all kinds of singularities have been removed.

The scalar field kinetic energies and potential energies have been represented towards the end of the work. For one parameter and three parameters models we found increasing nature of potential energy and decreasing nature of kinetic energy. On the contrary for two parameters model we observe decreasing nature in both potential energy as well as kinetic energy. Thus, the entire primordial universe expansion can be explained with a fluid system that is a mixture of the above models.

We have seen for one and three parameters model the scalar field potential obeys the slow-roll approximation theory. We have reconstructed the scale factor from the minimally coupled Friedmann equation that resolves initial singularity as well as finite time future singularity problems. Inflation occurs when $w_{EOS} \to -1$ the value of w is also in the negative region. The inflation will lie somewhere between w = -2/8 to -1/8. The nature of the potential energy curve is always increasing but its value becomes very high before inflation while it is low after inflation. None of the above-mentioned conditions are valid for two parameters model.

9. **Concluding remarks**

The Van-der-Waals fluid model is used in physics to introduce the idea of nonlinearity in system. A general equation of state is assumed to support this non-linearity. However, the EOS is also capable of providing an idea that is similar to the idea of dark energy from a fluid point of view. The equation of state for Chaplygin gas also represents the nonlinear system but is unable to provide this generalization. In this work we have tried to resolve this problem. Besides, we have provided the analysis of energy conditions and thermodynamics. The violation of SEC and positive differential entropy resolves the thermodynamic problems of dark energies. We have reconstructed the scale factor which is an additional result of our work that can discuss the phantom phase, Quintessence phase and post inflationary graceful exit phase in generalized way with changing the values of the parameters. So, we may conclude finally that our models and work are applicable in three primordial phases of cosmic evolution.

**Acknowledgement**

The authors Alokananda Kar and Shouvik Sadhukhan would like to thank Dr. Surajit Chattopadhyay for his help in this research and his guidance. The authors are thankful for the comments of the reviewers. The authors are thankful to the anonymous reviewer for the supportive comments. Financial support under the CSIR Grant No. 03(1420)/18/EMRII is thankfully acknowledged by Surajit Chattopadhyay

**Appendix**

**One parameter model:**

The one parameter model can be considered as follows.

$$P = \frac{8w\rho}{3-\rho} - 3\rho^2 \approx \frac{8w}{3}\rho + \left(\frac{8w}{9} - 3\right)\rho^2$$

(i)

Now considering $p = -\left(\frac{\partial U}{\partial V}\right)$ and $\rho = \frac{U}{V}$ we may write the following differential equation for internal energy. Here U = Internal energy and V = Volume scale factor.

$$\left(\frac{\partial U}{\partial V}\right) = -\frac{8w}{3}\left(\frac{U}{V}\right) + \left(3 - \frac{8w}{9}\right)\left(\frac{U}{V}\right)^2$$

(ii)

Now we can solve this partial differential equation to get the scale factor variation of internal energy.

**Two parameters model:**

The one parameter model can be considered as follows.

$$P = \frac{\gamma\rho}{1 - \frac{1}{3\rho_c}\rho} - \frac{9\gamma}{8\rho_c}\rho^2 \approx \gamma\rho - \frac{19\gamma}{24\rho_c}\rho^2$$

(iii)

Now considering $p = -\left(\frac{\partial U}{\partial V}\right)$ and $\rho = \frac{U}{V}$ we may write the following differential equation for internal energy. Here U = Internal energy and V = Volume scale factor.

$$\left(\frac{\partial U}{\partial V}\right) = -\gamma\left(\frac{U}{V}\right) + \frac{19\gamma}{24\rho_c}\left(\frac{U}{V}\right)^2$$

(iv)

Now we can solve this partial differential equation to get the scale factor variation of internal energy.

**Three parameters model:**

The one parameter model can be considered as follows.

$$P = \frac{\gamma\rho}{1 - \beta\rho} - \alpha\rho^2 \approx \gamma\rho + (\gamma\beta - \alpha)\rho^2$$

(v)

Now considering $p = -\left(\frac{\partial U}{\partial V}\right)_S$ and $\rho = \frac{U}{V}$ we may write the following differential equation for internal energy. Here U = Internal energy and V = Volume scale factor.

$$\left(\frac{\partial U}{\partial V}\right) = -\gamma\left(\frac{U}{V}\right) + (\alpha - \gamma\beta)\left(\frac{U}{V}\right)^2$$

(vi)

Now we can solve this partial differential equation to get the scale factor variation of internal energy.

According to the stability rules we may get,

**One parameter model:**

From Rule I:

$$\left(\frac{\partial p}{\partial V}\right)_S = -\left[\frac{24w}{(3-\rho)^2} - 6\rho\right] \frac{U_0(8w+3)^2(9+24w)V^{\frac{8w}{3}}}{3\left[(27-8w)+(9+24w)V^{\frac{8w+3}{3}}U_0\right]^2} < 0 \quad (vii)$$

From Rule II:

$$\left(\frac{\partial p}{\partial V}\right)_T = -\left[\frac{24w}{(3-\rho)^2} - 6\rho\right] \left[\left(\frac{(8w+3)ST}{U_0(9+24w)}\right)^{\frac{1}{2}}\left(\frac{8w+6}{6}\right)V^{-\frac{8w+12}{6}} - \frac{1}{2}\left(\frac{(8w+3)TS^{-1}V^{-\frac{(8w+6)}{3}}}{U_0(9+24w)}\right)^{\frac{1}{2}} \frac{(8w+3)(8w+6)(9+24w)U_0 V^{\frac{8w+9}{3}}}{3T\left[(27-8w)+(9+24w)V^{\frac{8w+3}{3}}U_0\right]^2} +$$

$$\frac{1}{2}\left(\frac{(8w+3)TS^{-1}V^{-\frac{(8w+6)}{3}}}{U_0(9+24w)}\right)^{\frac{1}{2}} \frac{2(8w+3)^2(9+24w)^2 U_0^2 V^{\frac{16w+12}{3}}}{T\left[(27-8w)+(9+24w)V^{\frac{8w+3}{3}}U_0\right]^3}\right] < 0 \quad (viii)$$

From Rule III:

$$c_V = T\left(\frac{\partial S}{\partial T}\right)_V = -\frac{(8w+3)(9+24w)V^{\frac{8w+6}{3}}}{\left[(27-8w)+(9+24w)V^{\frac{8w+3}{3}}U_0\right]^2} \frac{U_0}{T} < 0 \text{ (for positive value of } U_0) \quad (ix)$$

**Two parameters model:**

From Rule I:

$$\left(\frac{\partial p}{\partial V}\right)_S = -\left[\frac{\gamma}{\left(3-\frac{1}{3\rho_c}\rho\right)^2} - \frac{9\gamma}{4\rho_c}\rho\right] \frac{U_0(24)^2(\gamma+1)^3 \rho_c V^\gamma}{\left[19\gamma+24\rho_c(\gamma+1)V^{(\gamma+1)}U_0\right]^2} \quad (x)$$

From Rule II:

$$\left(\frac{\partial p}{\partial V}\right)_T = -\left[\frac{\gamma}{\left(3-\frac{1}{3\rho_c}\rho\right)^2} - \frac{9\gamma}{4\rho_c}\rho\right] \left[\left(\frac{\rho_c ST}{U_0}\right)^{\frac{1}{2}}\left(\frac{\gamma+2}{2}\right)V^{-\frac{\gamma+4}{2}} - \frac{1}{2}\left(\frac{\rho_c TS^{-1}V^{-(\gamma+2)}}{cU_0}\right)^{\frac{1}{2}} \frac{(24(\gamma+1))^2 \rho_c(\gamma+2)U_0 V^{\gamma+1}}{T\left[19\gamma+24\rho_c(\gamma+1)V^{(\gamma+1)}U_0\right]^2} +$$

$$\frac{1}{2}\left(\frac{\rho_c TS^{-1}V^{-(\gamma+2)}}{U_0}\right)^{\frac{1}{2}} \frac{2(24(\gamma+1))^3 \rho_c^2 U_0^2 V^{2\gamma+2}}{T\left[19\gamma+24\rho_c(\gamma+1)V^{(\gamma+1)}U_0\right]^3}\right]$$

$$(xi)$$

From Rule III:

$$c_V = T\left(\frac{\partial S}{\partial T}\right)_V = -\frac{(24(\gamma+1))^2 \rho_c V^{\gamma+2}}{\left[19\gamma+24\rho_c(\gamma+1)V^{(\gamma+1)}U_0\right]^2} \frac{U_0}{T} \quad (xii)$$

**Three parameters model:**

From Rule I:

$$\left(\frac{\partial p}{\partial V}\right)_S = -\left[\frac{\gamma}{(1-\beta\rho)^2} - 2\alpha\rho\right]\frac{U_0(\gamma+1)^3 V^\gamma}{[(\alpha-\beta\gamma)+(\gamma+1)V^{(\gamma+1)}U_0]^2} \tag{xiii}$$

From Rule II:

$$\left(\frac{\partial p}{\partial V}\right)_T = -\left[\frac{\gamma}{(1-\beta\rho)^2} - 2\alpha\rho\right]\left[\left(\frac{ST}{U_0}\right)^{\frac{1}{2}}\left(\frac{\gamma+2}{2}\right)V^{-\frac{\gamma+4}{2}} - \frac{1}{2}\left(\frac{TS^{-1}V^{-(\gamma+2)}}{U_0}\right)^{\frac{1}{2}}\frac{(\gamma+1)^2(\gamma+2)U_0 V^{\gamma+1}}{T[(\alpha-\beta\gamma)+(\gamma+1)V^{(\gamma+1)}U_0]^2} + \right.$$

$$\left.\frac{1}{2}\left(\frac{TS^{-1}V^{-(\gamma+2)}}{U_0}\right)^{\frac{1}{2}}\frac{(\gamma+1)^4 U_0^2 V^{2\gamma+2}}{T[(\alpha-\beta\gamma)+(\gamma+1)V^{(\gamma+1)}U_0]^3}\right] \tag{xiv}$$

From Rule III:

$$c_V = T\left(\frac{\partial S}{\partial T}\right)_V = -\frac{(\gamma+1)^2 V^{\gamma+2}}{[(\alpha-\beta\gamma)+(\gamma+1)V^{(\gamma+1)}U_0]^2}\frac{U_0}{T} \tag{xv}$$